\documentclass{an}
\usepackage{graphicx}
\usepackage{times}
\usepackage{fancyhdr}
\begin{document}
\title{Correction of SOHO CELIAS/SEM EUV Measurements Saturated by Extreme Solar Flare Events }

\author{L. V. Didkovsky, \inst{1}\
D. L. Judge, \inst{1}\ A. R. Jones, \inst{1}\ S. Wieman, \inst{1}\
B. T. Tsurutani, \inst{2} and D. McMullin, \inst{3}} \institute{
Space Sciences Center, Department of Physics and Astronomy,
University of Southern California, Los Angeles, California, U.S.A.
\and Jet Propulsion Laboratory, California Institute of Technology,
Pasadena, California, U.S.A. \and Praxis, Inc, Alexandria, Virginia,
U.S.A. }
\date{Received $<$date$>$;
accepted $<$date$>$; published online $<$date$>$}

\abstract {The solar irradiance in the Extreme Ultraviolet (EUV)
spectral bands has been observed with a 15~s cadence by the SOHO
Solar EUV Monitor (SEM) since 1995. During remarkably intense solar
flares the SEM EUV measurements are saturated in the central (zero)
order channel (0.1 -- 50.0~nm) by the flare soft X-ray and EUV flux.
The first order EUV channel (26 -- 34~nm) is not saturated by the
flare flux because of its limited bandwidth, but it is sensitive to
the arrival of Solar Energetic Particles (SEP). While both channels
detect nearly equal SEP fluxes, their contributions to the count
rate is sensibly negligible in the zero order channel but must be
accounted for and removed from the first channel count rate. SEP
contribution to the measured SEM signals usually follows the EUV
peak for the gradual solar flare events. Correcting the extreme
solar flare SEM EUV measurements may reveal currently unclear
relations between the flare magnitude, dynamics observed in
different EUV spectral bands, and the measured Earth atmosphere
response. A simple and effective correction technique based on
analysis of SEM count-rate profiles, GOES X-ray, and GOES proton
data has been developed and used for correcting EUV measurements for
the five extreme solar flare events of July 14, 2000, October 28,
November 2, November 4, 2003, and January 20, 2005. Although none of
the 2000 and 2003 flare $\it{peaks}$ were contaminated by the
presence of SEPs, the January 20, 2005 SEPs were unusually prompt
and contaminated the peak. The estimated accuracy of the correction
is about $\pm 7.5$\% for large X-class events.  } \correspondence{leonid@usc.edu}

\maketitle

\section{Introduction}
Solar Extreme Ultraviolet (EUV) variability has been observed by the
SOHO Solar EUV Monitor (SEM) (Judge et al. 1998) since Dec 1995. SEM
observations of extreme solar flare events with high temporal
resolution (15~s) may reveal still unknown relations between the
peak amplitude, the dynamics of a solar flare in the EUV bands, and
the resultant ionospheric effects (Meier et al. 2002; Tsurutani et
al. 2005). Tsurutani et al. (2005) compared some ``Halloween''
events (Oct-Nov, 2003) and the Bastille Day (BD, Jul 14, 2000) event
using SEM first order (Ch 1) 26 -- 34~nm count rates, the GOES x-ray
flux, the Libreville, Gabon Total Electron Content (TEC) data, and
the GUVI (TIMED) dayglow data. The comparison showed that the Oct
28, 2003 event was the most intense flare of the four analyzed
events in the SEM first order EUV wavelengths with the flare peak
amplitude about two times larger than the Bastille Day event. The
TEC increase was even larger, up to 25 TECU (Oct 28, 2003) compared
to about 5 -- 7~TECU for Nov 4, Oct 29 and the BD events.

Uncorrected SEM Ch 1 measurements of extreme solar flare events are
strongly contaminated after the SEP flux arrives. The time of
contamination depends on the SEP arrival time and may continue for
several hours. A model of the SEM response to a quasi-isotropic SEP
fluence allowed us to determine both the range of proton incident
energies and the function of SEM sensitivity to the SEP flux which
was found to have a maximum at about 12~MeV (Didkovsky et al. 2006).
The SEP count rate contribution to the Ch 1 signal may be
substantially larger than the EUV peak of an extreme solar flare.
This SEP contamination does not affect the measured EUV peak
amplitude of the solar flare when the SEP flux at 1 AU comes after
the EUV peak, but it creates a problem in the temporal analysis of
the flare if the particles arrive within 10 -- 15~min after the
beginning of the flare, as happened in the Jan 20, 2005 event
(Bieber et al. 2005).

The SEM zero order channel (Ch 0) detects both solar EUV and soft
X-ray irradiation in contrast with the 26 -- 34~nm Ch 1 spectral
band, and a comparison of the two could thereby show flare related
spectral variability. Unfortunately, the SEM Ch 0 measurements of
large X-class solar flares are saturated by the soft X-ray and EUV
fluxes during impulsive phases of these flares and must be
corrected.

Thus, correction of saturated Ch 0 and contaminated Ch 1 SEM
profiles is a critical task for analysis of EUV variability created
by the intense X-class solar flares which have a profound impact on
the Earth's atmosphere. Our first attempt to evaluate the amplitude
of the BD solar flare in the 19.5~nm bandpass was based on restored
saturated SOHO/EIT EUV images (Didkovsky et al. 2005). The flare
peak amplitude was modeled by extracting and integrating CCD pixel
signals from the area surrounding the flare assuming an efficient
pixel `bleeding' (blooming) process. If the efficiency of the
electron-hole charge transfer from the overfilled flare area to
neighboring pixels is high, integrating the electrons trapped in the
blooming areas of the EIT CCD can be used to rebuild the flare EUV
amplitude, and provide a lower limit to the solar flux.

Here we propose an advanced technique to correct saturated and
contaminated SEM measurements using Ch 0 and Ch 1 15~s count rates
together with GOES X-ray and GOES proton flux data. For the work
described in this paper, we used the term Ch 1 to reference a
combined, mean for the + and - first-order, measurements of the SEM.
Our model of SEM sensitivity to energetic particles (Didkovsky et
al. 2006) showed a peak sensitivity at about 12~MeV. It allows us to
use GOES proton data in the energy range of 8 -- 16~MeV as a
reference channel to represent the particle related increase in the
SEM Ch 1.

\section{Data Observations and Technique}

The present technique for correcting SEM saturated measurements was
applied to data from five extreme solar flare events: Jul 14, 2000,
Oct 28, Nov 2, and Nov 4, 2003, and Jan 20, 2005. In this paper we
will demonstrate a step by step procedure for correcting the BD SEM
measurements.

Figure 1 shows SEM high cadence count-rate data for the BD solar
flare event. The top panel represents the combined, mean for the `+'
and `-' first order, measurements (thick line) after subtracting the
pedestal of 272.4 cnt/s related to the full-disk solar background
flux. The thin line shows the GOES proton flux profile in the energy
range of 8 -- 16~MeV with the particle starting time of about 10.9
UT. The bottom panel shows the original (raw) Ch 0 SEM count rate
data saturated by the solar flare flux in the spectral band of 0.1
-- 50.0~nm.
\begin{figure}[h]
\resizebox{\hsize}{!}
{\includegraphics[width=20pc]{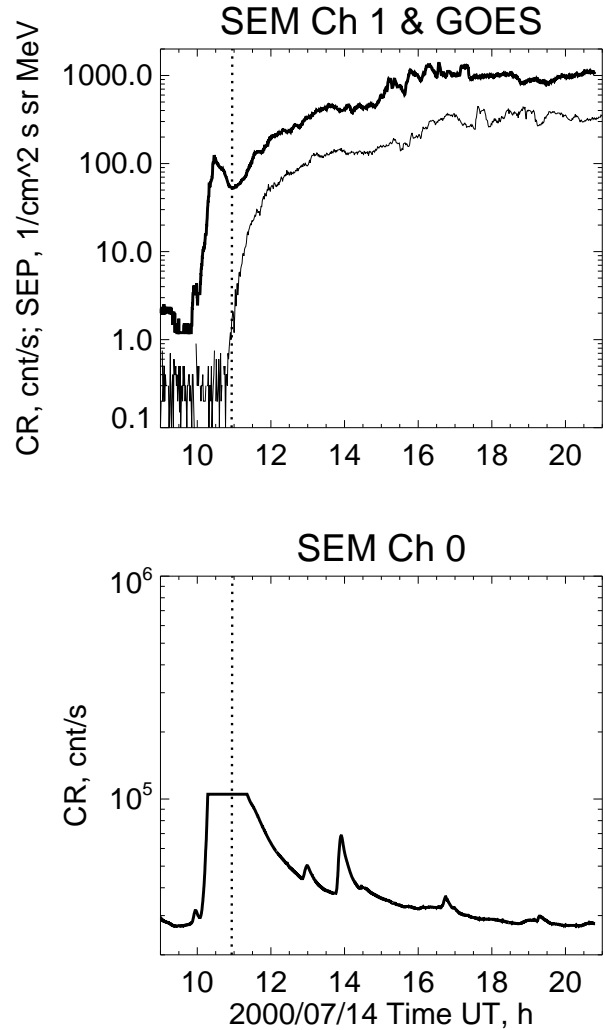}} \caption{SEM
count-rate (CR) measurements (thick lines) and GOES 8 -- 16~MeV
 particle flux (thin line) for the BD solar flare event. The SEM Ch 1 (top panel) is
 contaminated by the rise SEP flux after the time indicated by the vertical dotted line.
 The SEM Ch 0 (bottom panel) is saturated
 by the soft x-ray and EUV flux. The unsaturated decay portion of the SEM Ch 0 profile is only weakly
 affected by the increased SEP flux. }
\end{figure}

\subsection{Assumptions and Simplifications of the Method}

A 100\% accurate correction of the saturated SEM Ch 0 is not
possible, even for a known EUV response function of the SEM, due to
the unknown spectral energy distribution of the solar flare in the
observed spectral region. The unknown particle energy-flux
distribution at the SOHO location presents a similar problem for the
correction of the solar flare associated SEP-contamination of the Ch
1 signal. Thus, we have to make some simplifications and apply some
assumptions resulting in corrected profiles that are of a lower
level of accuracy than the typical data, (absence of extreme solar
flares) which is both unsaturated and uncontaminated.

The simplification, which affects the amplitude of the saturated Ch
0 electromagnetic flux measurements is the replacement of the real
(unknown) shape of the flare peak in the spectral band of 0.1 --
50~nm with a modeled peak based on two converging power-law curves.
Usually, this leads to a sharper (larger amplitude) modeled peak
than the smoother real peak, introducing a peak amplitude error of a
few percent. Each of the two converging power-law curves is modeled
using a large number of count rate points (20 -- 60 or more) on the
lower amplitude channel profiles, either in the initial portion of
the flare rise time profile or in the post-saturated decay profile.
In addition to modeling the power-law curves with a large number of
count rate points in the low amplitude parts of the flare profiles,
we note that the amplitude of the peak restored with two power law
curves depends on the temporal position of the converging point.
Comparing the peak position in other spectral bands, we have found
that with a few minutes uncertainty, the temporal position of the Ch
0 peak may be considered between positions of the GOES XL and the
SEM Ch 1 peaks.

Another simplification, which does not affect the flare calculated
peak amplitude and has a minor impact on the ionospheric effects, is
the modeling of a flare decay profile. The Ch 1 uncorrected flare
decay profile is usually contaminated by the SEP signal but,
typically, well after the peak of the EUV flux. In contrast, the Ch
0 count rate signal for the decay profile of a solar flare has a
substantially larger component related to the EUV and soft X-ray
irradiation than to the SEP flux. With a spectral similarity
simplification, the measured Ch 0 EUV and soft X-ray decay profile
may be transferred with an appropriate scaling to the corrected Ch 1
profile.

 \subsection{Step by Step Procedure of Correction}

A first step in the correction procedure includes two preparatory
operations: a ``clean up'' of the Ch 0 decay profile by subtracting
the SEP related count rate signal, and determination of the temporal
position of the flare peak point. The Ch 0 decay profile consists of
a sum of unknown X-ray, EUV and SEP related portions. With an
assumption of a quasi isotropic SEP flux and similar Ch 0 and Ch1
instrumental response to the protons (Didkovsky et al. 2006), the
SEP related portion of the Ch 0 signal may be considered as known
from either the GOES (8 -- 16~MeV) channel or from the Ch 1 count
rate data. For extreme solar flare events with intense SEP flux, Ch
1 count rate measurements for the decay portion of the EUV solar
flare profile are strongly contaminated by the SEP flux, with a SEP
to EUV ratio of about 100 to 1. This means that Ch 1 data for the
decay phase of the flare may be considered as a SEP signal with an
error of about 1 \%. Then, the EUV Ch 0 count rate signal for the
decay phase, $EY0_{D}$, may be modeled as
\begin{equation}
 EY0_{D}(k)= Y0_{D}(k)-Y0_{SEP}(k)
 \end{equation}
where $Y0_{SEP}$ is the SEP signal determined from the
 contaminated Ch 1 measurements as
 \begin{equation}
 Y0_{SEP}(k)=Y1_{D}(k)
 \end{equation}
 where $k$ is the index which corresponds to the time when the SEPs started
 to arrive and $Y1_{D}$ is the Ch 1 measurement. After this
 procedure a portion of the Ch 0 unsaturated decay profile consists
 of (with accuracy of about 99 \%) the X-ray and EUV signal and thus may be used
 to model the whole decay profile with a power law
 curve in the next step.

An evaluation of the time when the peak of the flare in the Ch 0
spectral band occurs is based on a comparison of temporal position
of unsaturated Ch 1 and GOES XL (0.1 -- 0.8~nm) flare peaks. The
GOES channel XL has the same short wavelength limit as the SEM Ch 0
(0.1 -- 50~nm) and represents the input from the soft X-ray
radiation. The SEM Ch 1 (26 -- 34~nm) adds the spectral irradiation
component in the middle of the Ch 0 spectral band. A comparison of
the temporal position of the unsaturated Ch 0 peak for a number of
X-1 class solar flares (not in this paper) showed that maximal Ch 0
fluxes occur quite close (no more than a few minutes difference) to
the time when the GOES XL channel had measured the corresponding
flare peaks. For the BD solar flare we compared (Figure 2) both
GOES-8 XL and XS (0.05 -- 0.3~nm) flare profiles (thin lines) and
the position of the flare peak for the Ch 1 signal (dotted line).
\begin{figure}[h]
\resizebox{\hsize}{!}
{\includegraphics[width=18pc]{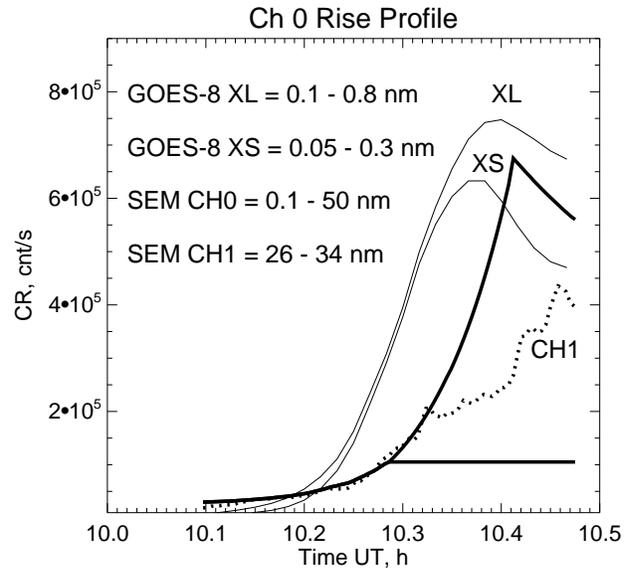}}
 \caption{A comparison of the temporal position of the solar flare peak in both GOES-8
 X-ray channels (XL and XS) and the SEM Ch 1 (dotted line) for the BD 2000 solar flare.
 The location
 of the peak for the Ch 0 (thick line) is assumed to be between the Ch 1 and
 XL peaks. Ch 1 profile (dotted line) was scaled up to fit the initial rise phase of
 the Ch 0 profile. }
\end{figure}
 We have assumed that the temporal position of the Ch 0 peak should be between the GOES-8 XL peak and the
 Ch 1 peak. The exact temporal position of the flare peak in the
 SEM Ch 0 is unknown. With the above assumption the uncertainty of the time of the Ch 0 peak is about $\pm 1.8$~min,
 which leads to a peak amplitude uncertainty of about $\pm 5.5\%$.

The second step in the correction is modeling of the whole decay
portion of the saturated Ch 0 profile. The power law curve should
fit the corrected EUV portion of the Ch 0 profile and be extended to
the time of the evaluated peak of the flare.To model this profile we
fit the ``cleaned up'' EUV signal in the first step, $EY0_{D}(j)$,
to a j-portion of the power law curve $Y_{D}(i)$, minimizing the
distance between them:
\begin{equation}
 \sum_{j}(Y_{D}(j)-EY0_{D}(j))^{2} = min
 \end{equation}
 where $Y_{D}(i)$ is
\begin{equation}
 Y_{D}(i) = \frac{a}{(i-b)^{k}} + P_{0}
 \end{equation}
and $a$, $b$, and $k$ are constant coefficients, $i$ is the index,
which begins with the time of the flare peak, $j$ is the index, $j <
i$, related to the unsaturated decay portion of the Ch 0 profile,
and $P_{0}$ is the Ch 0 pedestal. The important point here is to fit
the modeled curve to the measured curve after `removing' the signals
associated with both particles and EUV fluctuations. This means that
the entire modeled curve (3) for the decay profile should `ignore'
both EUV and SEP fluctuations.

Figure 3 shows the modeled (dotted line) fit of the decay signal
profile. The identification of the Ch 0 EUV fluctuations is done
using the GOES XL reference signal shown in Figure 3 with the thin
line.
\begin{figure}[h]
\resizebox{\hsize}{!}
{\includegraphics[width=18pc]{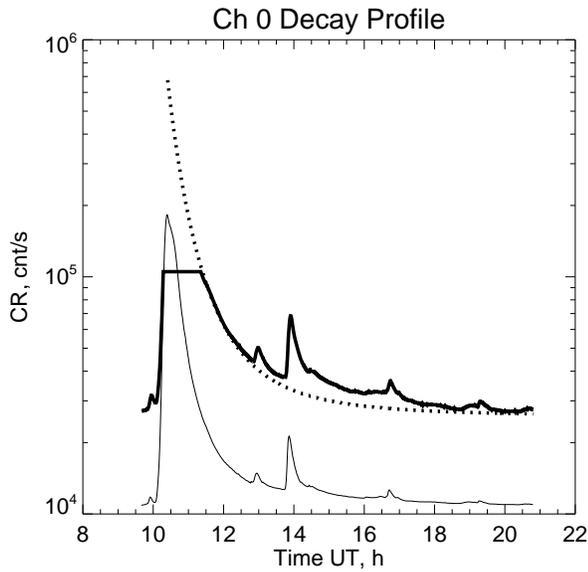}}
 \caption{A modeled curve (dotted line) to fit the decay signal profile of the BD 2000 extreme solar flare.
 Some signal
 fluctuations between about 13 and 20 UT are due to the EUV flux fluctuations visible in the GOES
 reference XL (soft X-ray) profile (thin line).}
\end{figure}
The entire modeled curve (4) is used to determine the amplitude of
the peak and the shape of the decay profile.

The third step in modeling is to fit the rise Ch 0 profile using the
initial (unsaturated) portion of the Ch 0 profile and the amplitude
of the peak determined in the second step. Figure 2 (above) shows
the fitting curve of the rise profile. The Ch 1 measurements
$Y1_{R}$ are scaled up ($Y1_{R}*n1$) to fit the initial rise time
portion of the Ch 0 profile (before the saturation), shown by the
dotted line in Fig 2. We have analyzed a number of X-class solar
flares (not the issue of this work) to see how well the scaled Ch 1
could fit Ch 0 and found that they usually fit each other within 5
-- 7\%. This fitting ratio is used in the next step to scale back
the Ch 0 decay profile to the Ch 1 decay profile for complete
correction of the channel.

The final step is to scale back the Ch 0 corrected decay profile to
the Ch 1:
 \begin{equation}
 Y1_{B}(j)=Y0_{B}(j)/n1
 \end{equation}
 The scaled back Ch 0 profile should match the entire decay profile of
 the Ch 1 at the point where Ch 1 is contaminated by the SEPs.
 The corrected profiles of both channels are shown in Figure 4.
\begin{figure}[h]
\resizebox{\hsize}{!}
{\includegraphics[width=18pc]{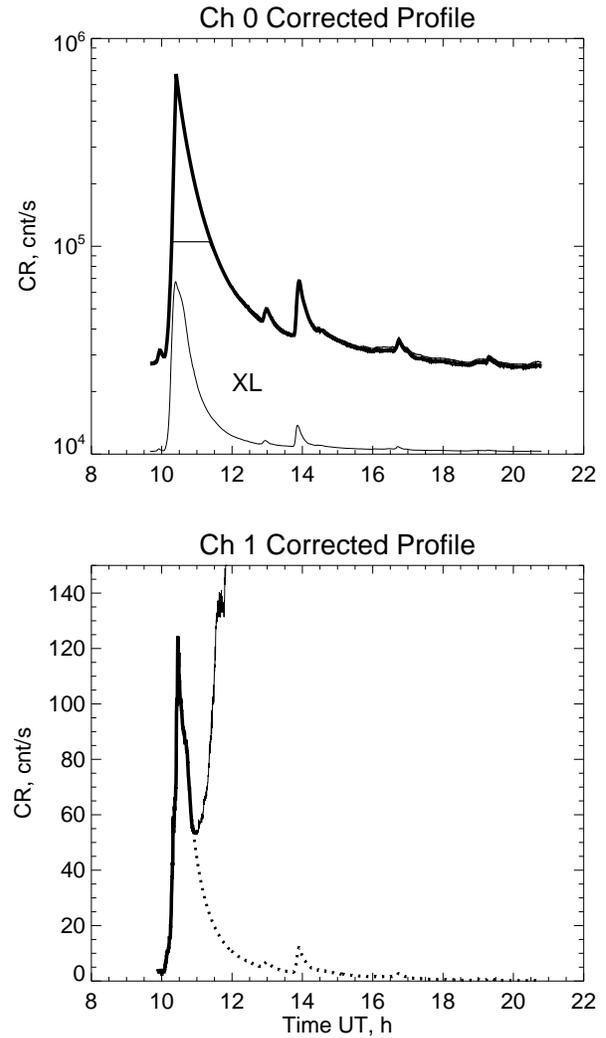}}
 \caption{Restored profiles of the Ch 0 (top) and the Ch 1
 (bottom) count rates. The thin line on the top portion of Fig 4 shows
 the GOES XL profile as a reference for the EUV related fluctuations on the
 corrected decay parts of the flare. }
\end{figure}

\section{Results}

The results of correction of the SEM data for the analyzed events
are shown in Table 1.
\begin{table}[h]
\caption{Summary of corrected amplitudes for the analyzed extreme
solar flares}
\begin{tabular}{lccc}
\hline
\multicolumn{1}{c}{Date}&Ch 1, cnt/s & Ch 0, 1E6 cnt/s & Time interval\\
& & & between Ch 1\\
& & & and XL \\
& & & peaks, min \\
\hline
Jul 14, 2000 & $124.4 \pm 2.4$ & $0.68 \pm 0.075$ & 3.6\\
Oct 28, 2003 & $218.6 \pm 4.4$ & $1.55 \pm 0.30$  & 4.8\\
Nov 02, 2003 & $44.7  \pm 0.9$ & $0.68 \pm 0.18$  & 12\\
Nov 04, 2003 & $105.0 \pm 2.1$ & $1.42 \pm 0.07$  & 3.6\\
Jan 20, 2005 & $202.8 \pm 12.6$ & $1.40 \pm 0.20$  & 3.0\\
\hline
\end{tabular}
\end{table}
Note that none of the 2000 -- 2003 flare peaks were contaminated by
the SEP fluxes and that the EUV Ch 1 peak amplitudes were determined
directly from the SEM measurements. The errors of about $\pm 2.0\%$
for Ch 1 (and less for Ch 0) are related to the method of
subtracting the full disk solar background. The Jan 20, 2005 EUV
flare peak was contaminated by SEPs and the error of about $\pm
7.0\%$ is related to both the solar disk background, and to some
uncertainty in the SEM Ch 1 response to the SEP flux. The peak
amplitude mean error related to the uncertainty of the Ch 0 peak
position is about $\pm 7.5\%$ for the analyzed extreme solar flare
events.

\section{Concluding Comments}

We have developed a method for correcting SEM saturated Ch 0 and
contaminated Ch 1 count rate measurements for five extreme solar
flare events. The correction algorithm is based on SEM EUV data
(both channels), GOES XL, and GOES proton flux measurements.

Comparing the temporal position of a usually unsaturated SEM Ch 1
flare peak with an unsaturated GOES XL peak gives us the temporal
range for modeling the Ch 0 flare peak. The length of this temporal
range (see Table 1, last column) is proportional to the uncertainty
of the Ch 0 peak position. Calculating the Ch 0 peak amplitude at
the limits of this interval allowed us to estimate the mean
uncertainty of the Ch 0 corrected amplitude to be about $\pm 7.5\%$.

SEM modeled sensitivity to SEPs allowed us to use the GOES proton
flux measurements in the energy range of 8 -- 16~MeV as a temporal
reference. These measurements show the approximate start time of Ch
1 contamination by the SEPs. This information was used to correct
the Ch 1 flare profile by subtracting the SEP count rate from the Ch
1 data. After correcting the Ch 1 profile, it is possible to
determine the net SEP flux in the energy range of the modeled SEM
sensitivity.

In the correction process a flare specific Ch 0 to Ch 1 count rate
normalization ratio, $n1$, was determined to fit the Ch 1 count rate
data profile prior to Ch 0 saturation. Following Ch 0 saturation we
assume that the Ch 1 count rate decay is proportional to the Ch 0
count rate, and with the same $n1$ scaling factor.

Restored SEM EUV Ch 0 flare peak amplitudes for large (X-class)
solar flare events allow us to extend the previous (Tsurutani et al.
2005) study of the effects of the Ch 1 and Ch 0 EUV flare dynamics
and the resulting effects on the ionosphere and the Earth's
atmosphere generally. The proposed algorithm allows one to model the
entire Ch 1 flare profile from beginning to end to obtain the total
26-34 nm flux.  This could be useful for detailed ionospheric
modeling work in the future.
%
%

\begin{acknowledgements}
This work was supported in part by the SURP, a Partnership Between
JPL and USC (JPL Subcontract 1278368), and by NASA grants NNG05WC09G
and 153-5979. Portions of this research were performed at the Jet
Propulsion Laboratory, California Institute of Technology, under
contract with NASA.

\end{acknowledgements}

\end{document}